\documentclass[sigconf]{acmart}

\AtBeginDocument{%
  }
\usepackage{graphicx}
\usepackage{framed}
\usepackage{xcolor}
\definecolor{shadecolor1}{rgb}{0.92, 0.86, 0.89} 
\definecolor{shadecolor2}{rgb}{1, 0.99, 0.81} 
\usepackage{listings}
\usepackage{makecell}
\setcellgapes{2pt} 
\makegapedcells
\usepackage{subcaption}
\newcommand{\shortname}{\textsl{LELANTE}} 
\setcopyright{acmlicensed}
\copyrightyear{2025}
\acmYear{2025}
\acmDOI{XXXXXXX.XXXXXXX}
\acmConference[EASE 2025]{The 29th International Conference on Evaluation and Assessment in Software Engineering, June 17–20, 2025, Istanbul, Turkey}
\acmISBN{978-1-4503-XXXX-X/2025/06}

\begin{document}

\title{LELANTE: LEveraging LLM for Automated ANdroid TEsting}

\author{
Shamit Fatin\textsuperscript{1}, Mehbubul Hasan Al-Quvi\textsuperscript{1},
Sukarna Barua\textsuperscript{1}, Anindya Iqbal\textsuperscript{1}, 
Sadia Sharmin\textsuperscript{1}, Md.\ Mostofa Akbar\textsuperscript{1},
Haz Sameen Shahgir\textsuperscript{2},
Kallol Kumar Pal\textsuperscript{3},
A.\ Asif Al Rashid\textsuperscript{3}
}
\affiliation{%
  \institution{\textsuperscript{1}Bangladesh University of Engineering and Technology, Bangladesh}
  \country{}
}
\affiliation{%
  \institution{\textsuperscript{2}University of California, Riverside, USA}
  \country{}
}
\affiliation{%
  \institution{\textsuperscript{3}Samsung R\&D Institute Bangladesh, Bangladesh}
  \country{}
}
\email{
    {1805055,1805007,sukarnabarua,anindya,sadiasharmin,mostofa}@cse.buet.ac.bd
}
\email{
    hshah057@ucr.edu
}
\email{
    {kallol.kumar,ahmad.asif}@samsung.com
}


\begin{abstract}
    Given natural language test case description for an Android application, existing testing approaches require developers to manually write scripts using tools such as Appium and Espresso to execute the corresponding test case. This process is labor-intensive and demands significant effort to maintain as UI interfaces evolve throughout development. In this work, we introduce \shortname{}, a novel framework that utilizes large language models (LLMs) to automate test case execution without requiring pre-written scripts. \shortname{}  interprets natural language test case descriptions, iteratively generate action plans, and perform the actions directly on the Android screen using its GUI. \shortname{} employs a screen refinement process to enhance LLM interpretability, constructs a structured prompt for LLMs, and implements an action generation mechanism based on chain-of-thought reasoning of LLMs. To further reduce computational cost and enhance scalability, \shortname{} utilizes model distillation using a foundational LLM. In experiments across 390 test cases spanning 10 popular Android applications, \shortname{} achieved a \textbf{73\%} test execution success rate.  Our results demonstrate that LLMs can effectively bridge the gap between natural language test case description and automated execution, making mobile testing more scalable and adaptable.

\end{abstract}

\ccsdesc[500]{Software and its engineering~Software testing and debugging}
\ccsdesc[100]{Computing methodologies~Planning for deterministic actions}
\ccsdesc[100]{Computing methodologies~Natural language processing}

\keywords{Automated Android Testing, Large Language Models (LLMs), GUI Refinement, Model Distillation, Prompt Engineering}

  


\makeatletter
\def\@shortauthors{Shamit, et al.}
\hypersetup{pdfauthor={Shamit, et al.}}
\makeatother
\maketitle

\begin{figure}[H]
    \centering
    \includegraphics[width=\linewidth]{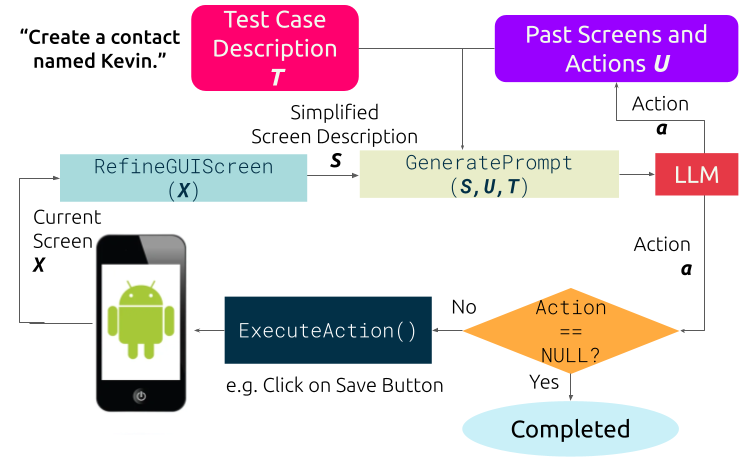}
    \caption{Overview of the \shortname{} test case execution process. Given a natural language test case description, \shortname{} incrementally refines the screen GUI representation and executes the test case step by step until completion.}
    \label{fig:execution_pipeline}
\end{figure}
\vspace{-\baselineskip}
\section{Introduction}

The rapid growth of the mobile application ecosystem has made Android the dominant platform for app development, with a market share consistently exceeding 70\% \cite{rashidi2015survey}. This diversity allows for innovation but also presents challenges for developers aiming to ensure consistent functionality across various devices, operating system versions, and screen sizes \cite{sarkar2019android, jaiswal2018android, dini2018risk, 10.1145/1814433.1814453}. As mobile applications evolve, rigorous testing becomes essential to maintaining software quality. 

 
Application testing involves designing and executing \textbf{test cases}, which define scenarios to validate an application’s functionality. Each test case typically includes a \textbf{test case description}, outlining the overall objective of the test. For example, a test case may verify whether a user can successfully log into an application, and its description would be \texttt{"Log in the application by entering user name and password"}. In some cases, a test case may also specify a sequence of \textbf{test steps} or \textbf{actions}, which are atomic, human-performable interactions (e.g., \texttt{tapping a button}, \texttt{entering text}) that must be executed in order to complete the test case. For instance, the test steps for the login test case might include \texttt{launching the application}, \texttt{tapping the "Login" button}, \texttt{entering a valid email and password}, \texttt{pressing "Sign In"} and \texttt{verifying that the home screen appears}. Both test case descriptions and test steps are written in human-readable language and are not structured for direct machine interpretation.

\textbf{Quality Assurance (QA) teams} are responsible for creating test cases to ensure comprehensive coverage of an application’s features and expected behavior. Human testers then execute these test cases by either following the specified test steps or, if no steps are provided, relying on their own understanding of the test objective. However, manually defining test steps is a labor-intensive process that requires continuous updates as applications evolve. This challenge is particularly pronounced in large development teams, where testing team may be unfamiliar with the application’s functionality. In this work, we assume that test steps are not explicitly provided by the QA team. Therefore, the system must autonomously determine the appropriate sequence of actions to execute each test case.


Automated testing frameworks such as Appium \cite{AppiumDocs} and Espresso \cite{EspressoDocs} allow testers to write scripts for test cases that simulate \textbf{test steps} on an application. Although these tools ensure consistency and repeatability, they require significant manual effort and technical knowledge to create, maintain, and rewrite test scripts, especially as user interfaces evolve throughout development \cite{kong2018automated, lin2020test, said2020gui, coppola2020mobile, grano2018exploring}. Record-and-replay mechanisms \cite{sahin2019randr}, which capture user actions for later execution, simplify the scripting process to some extent but struggle with dynamic UI modifications \cite{guo2019sara}. These limitations necessitate an automation tool that can directly interpret pre-existing human-readable test cases description in order to execute the test case without requiring manual intervention.

Recent advances in artificial intelligence, particularly in large language models (LLMs) such as GPT-4 \cite{achiam2023gpt} and LLaMA \cite{touvron2023llama}, have enabled new possibilities for automating human-readable tasks\cite{liu2024your, hou2024large, zimmermann2023gui, li2024test, konuk2024evaluation}. For example, AutoDroid \cite{wen2024autodroid} employs LLMs to automate Android tasks, demonstrating feasibility in interpreting and executing simple user commands. However, AutoDroid is not designed for structured test case execution, as it lacks mechanisms for dynamically adapting to UI changes and recovering from execution errors and plan for knowledge distillation. Furthermore it relies on foundational LLMs without incorporating model distillation techniques, making its execution computationally expensive and less practical for large-scale testing. These limitations make it unsuitable for comprehensive and scalable test automation. 

In this work, we introduce \textbf{\shortname{}}, a novel framework that automates test case execution using LLMs, given only a natural language test case description. \shortname{} eliminates the need for manually written test scripts by automatically generating a test case execution plan, consisting of required action sequence for executing the test case. \shortname{} guides an LLM to generate actions iteratively using a structured prompt based on a refined screen representation, past action history, and test case description. The generated actions are executed on the application until \shortname{} determines that the test case execution is complete. Figure~\ref{fig:execution_pipeline} illustrates the core execution process of \shortname{}. By automating this entire process, \shortname{} significantly reduces the human effort required for software testing, allowing teams to conduct more extensive testing with minimal overhead. ~\shortname{} eliminates the need for writing detailed and time-consuming test case steps, as it executes the test case directly from test case descriptions.
%

Through experiments across 390 test cases spanning 10 popular Android applications, \shortname{} achieved a \textbf{73\% test execution success rate}. Our findings demonstrate that LLM-based execution significantly improves the adaptability and scalability of mobile testing workflows, reducing the reliance on brittle, manually maintained scripts. This paper presents the following key contributions:
\begin{itemize}
    \item A fully automated test case execution framework that eliminates the need for manually written scripts, allowing natural language \textbf{test case descriptions} to be executed directly and dynamically on the Android application.
    \item A structured prompt to guide the LLM and a GUI refinement process to extract relevant interactive elements from screen XML representations, reducing noise and improving the LLM's ability to accurately interpret the application interface and reason actions.
    \item Efficient distillation of open-source large language models for \shortname{} execution, reducing computational overhead while maintaining accuracy.
    \item Empirical validation demonstrating a \textbf{73\% test execution success rate} across \textbf{390 test cases}, showcasing the effectiveness of LLM-driven automation.
\end{itemize}

\section{Problem Statement}

Let $T$ denote a natural language test case description, and $S$ represent the state of the Android application system at any given time. The state $S$ includes all observable and internal aspects of the system, such as the current screen layout, UI components, system settings, network conditions, and application runtime states. From $S$, we obtain a simplified and structured representation $G$, which captures only essential components of the state including UI elements and their properties, allowing for reasoning about available interactions.

At the start of testing, the Android device is in an initial state $S_0$. The execution of an action or a test step $a_t$ causes a transition in the system, leading to a new state $S_t$, modeled by the state transition function $\Phi$:
\[
S_t = \Phi(S_{t-1}, a_t).
\]
The function $\Phi$ captures how Android updates its state in response to user interactions. For instance, if the action $a_t$ is "tap on a button," $\Phi$ updates $S_{t-1}$ to reflect the resulting UI change, such as navigating to a new screen or displaying a pop-up dialog. Similarly, if $a_t$ is "enter text in a field," $\Phi$ modifies $S_{t-1}$ to include the new text input.

The test case description $T$ implicitly defines an expected final state $S_e$, which specifies the conditions that must be satisfied for the test to be considered successful. For example, if $T$ is \texttt{"Delete the contact named X"}, then $S_0$ represents the initial state where the contact exists in storage. The sequence of actions $\{a_1, a_2, \dots, a_n\}$ may involve opening the contacts app, searching for the contact, selecting the delete option, and confirming the deletion. The expected final state $S_e$ should then satisfy the condition that the contact is no longer retrievable in the storage system.

The goal of \shortname{} is to generate a valid sequence of actions $\{a_1, a_2, \dots, a_n\}$ such that the android application transitions from $S_0$ to $S_e$ in compliance with the test case. At each step $t$, \shortname{} acts as an agent that determines the next action:
\[
a_t = f\bigl(T, \{a_1, a_2, \dots, a_{t-1}\}, G\bigr),
\]
ensuring that after $n$ steps, the system reaches $S_n = S_e$. \shortname{} employs iterative decision-making to automate Android testing to continuously adapt its actions based on the observed UI state.

A test case is considered successfully completed if, after executing a valid sequence of actions \( \{a_1, a_2, ..., a_n\} \), the Android application's final state \(S_n\) matches the expected final state \(S_e\) implied by the test case description \(T\). If the final state does not satisfy the conditions implied by the description, the test case is considered incomplete or failed.

\begin{figure*}
    \centering
    \includegraphics[width=0.8\linewidth]{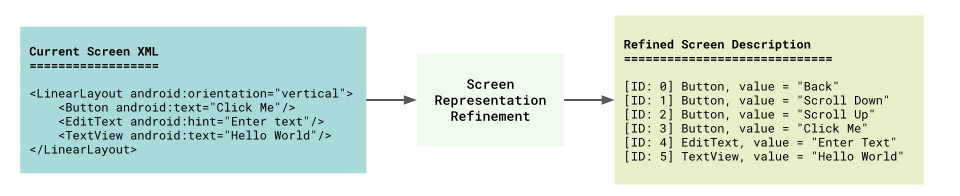}
    \caption{Screen representation refinement for LLM understanding. This process extracts essential and actionable elements (e.g., buttons, text fields) from the original screen representation  (XML) and convert it to an optimized format suitable for LLM prompts.}
    \label{fig:test_execution_xml_simplification}
    \Description {-}
\end{figure*}

\section{Methodology} \label{sec:methodology}


\shortname{} begins with refining the application's graphical representation, eliminating unnecessary metadata while preserving only relevant interactive elements to create a structured and meaningful screen description (\S~\ref{sec:gui_refinement}). This refined representation is then used to generate a structured prompt (\S~\ref{sec:llm_execution}), enabling the LLM to determine the next action based on the provided test case description and execution history. Once an action is selected, it is mapped to the corresponding UI element and executed on the device, with built-in mechanisms for ambiguity resolution and iterative GUI interaction (\S~\ref{sec:action_execution}). To ensure cost-effective execution without reliance on third-party vendors, \shortname{} employs model distillation using an open-source large language model, significantly reducing computational overhead while maintaining accuracy (\S~\ref{sec:model_distillation}).

\vspace{-0.5\baselineskip}
\subsection{Screen Representation Refinement} \label{sec:gui_refinement}
Android applications store screen layouts as XML structures, which, while comprehensive, are often verbose and inefficient for direct understanding and processing by LLMs. \shortname{} applies a preprocessing step to refine the screen representation by extracting only the essential interactive and informative components from an XML screen description. The refinement process begins with XML parsing, removing elements that lack interaction potential, such as non-essential and non-interactive background images, labels, and decorative UI components. Interactive elements, including buttons, text fields, and checkable inputs, are retained. Each actionable element is assigned with an unique identifiers to enable referencing during execution. Some retained non-interactive elements include textual descriptions or icons. Since \shortname{} relies on a purely text-based LLM for action inference, an image-to-text icon classification model is utilized to convert icon images into textual descriptions. Based on a Vision Transformer (ViT) model \cite{dosovitskiy2021an}, the icon-classifier is integrated in~\shortname{} to recognize icons and include their semantic labels in the representation. 

In addition to existing screen elements,~\shortname{} explicitly adds navigation buttons for scrolling and a 'Back' action, to the refined screen representation. Inclusion of the 'Back' button provides a built-in recovery action, enabling~\shortname{} to backtrack (guided by reasoning from LLM)  and rectify erroneous action sequences autonomously.  Figure \ref{fig:test_execution_xml_simplification} provides an example of this refinement process.

\vspace{-0.5\baselineskip}
\subsection{LLM-Driven Action Generation} \label{sec:llm_execution}



Given natural language test case description, refined screen representation, and a record of previous actions,~\shortname{} constructs a structured prompt to guide the LLM in reasoning about the next action. The prompt is designed to enforce a structured chain-of-thought \cite{wei2022chain} reasoning process, ensuring that the model first thinks about how to reach the final state before generating appropriate action. The overall structure of the prompt is given in Appendix~\ref{app:prompt}. 

A key challenge in this process was resolving ambiguity when the LLM selected an action. To address this, \shortname{} assigns unique identifiers to each actionable UI element during the screen refinement stage, ensuring that the LLM can reference them precisely. Instead of generating free-text descriptions of UI elements, the LLM directly outputs the corresponding identifier, eliminating potential misinterpretations and reducing the need for additional disambiguation steps.

An important component of our structured reasoning mechanism is error recovery. During the execution, \shortname{} analyzes the execution history and current screen to autonomously detect whether it has executed any incorrect or unintended action through iterative reasoning steps. Upon detection of an erroneous step, \shortname{} chooses the explicitly provided 'Back' button in the refined screen representation to revert to an earlier screen state, from which it re-evaluates and selects an alternative action in subsequent steps to rectify its action plan. The reasoning process repeats the backtracking until a viable action path is found or until the test case is successfully completed.

To programmatically determine when a test case execution should end, \shortname{} explicitly instructs the LLM, within the structured prompt described in Appendix~\ref{app:prompt}, to reason about whether the expected final state described in problem statement has been reached. Specifically, the LLM sets a boolean indicator when it detects completion, accompanied by a unique action identifier \texttt{-1}. \shortname{} detects this special identifier, signaling termination of the ongoing test execution. Finally, a human tester manually validates if the actual application state matches the intended outcome from the test case description, confirming successful execution.

\vspace{-0.5\baselineskip}
\subsection{Action Execution Using GUI Interaction} \label{sec:action_execution}


After generating an action by LLM,~\shortname{} executes it on the application GUI using the identified UI element and applying the interaction (e.g, input a text or click a button). We use Appium to execute LLM provided action into the Android GUI.  Once the action is executed, the application state changes and the XML representation of the new screen is fetched by Appium. This screen along with previous actions and test steps are iteratively processed through the~\shortname{} pipeline until it decides that no further action is required.



\vspace{-0.5\baselineskip}
\subsection{Distillation for Cost-Effective Execution} \label{sec:model_distillation}

While foundational LLMs (e.g., GPT-4o and Claude) demonstrate strong reasoning capabilities, their computational demands make them impractical for local or cheap test execution. To address this, \shortname{} employs knowledge distillation~\cite{gou2021knowledge}, where a smaller model is fine-tuned using data generated by a larger model. This reduces computational overhead while preserving structured reasoning capabilities.

The distillation process began with human testers executing test cases using a foundational model (GPT-4o), during which we recorded its responses, including predicted actions and reasoning steps. We collected test cases for \textbf{30 Android applications} that were not included in the final evaluation set. To maintain high-quality training data, we filtered out erroneous actions, redundant backtracking, and inefficient execution paths, resulting in a fine-tuning dataset consisting of 4,000 test case steps.

For model adaptation, we employed \textbf{Low-Rank Adaptation (LoRA)}~\cite{hu2022lora}, a parameter-efficient fine-tuning approach, allowing us to optimize the model on local machines without excessive computational overhead. Because we employ a step by step execution model, we needed to train the model on the test steps rather than in full execution history. Our fine-tuned model retained the structured reasoning of the foundational model while significantly reducing inference costs, making~\shortname{} more practical for real-world deployment. 

\section{Evaluation}

We evaluate \shortname{} based on its accuracy and efficiency in executing natural language test cases, answering the following two research questions:

\textbf{RQ1: How accurately and efficiently can \shortname{} execute natural language test cases across various applications?}

\textbf{RQ2: How effectively can LELANTE recover from erroneous steps?}

Our evaluation is performed using a dataset of \textit{390 human-generated test cases} for \textit{10 Android applications}. The minimum number of test steps required to complete each test case varied. Tests were conducted on physical Android devices (Samsung A14, A15, and A03 models) running Android 11, replicating realistic testing environments.

\paragraph{Evaluation metrics.} We use the following three metrics:
\begin{itemize}
    \item \textbf{Test Execution Success Rate}: Percentage of test cases completely executed.
    \item \textbf{Error Recovery Rate}: Percentage of erroneous steps after which \shortname{} autonomously detects mistakes, navigates backward, and chooses the correct alternative step.
    \item \textbf{Execution Time per Step}: Average time required to perform each test step.
\end{itemize}

\paragraph{Performance of LELANTE in test case execution.}
The evaluation results are summarized in Table~\ref{tab:rq1_results}. For RQ1, the \textbf{73\% success rate} demonstrates \shortname{}'s reliability and efficiency in executing natural language test cases description without predefined scripts. For RQ2, the \textbf{78\% error recovery rate} highlights the framework’s capability to dynamically handle execution errors. Due to limited prior work in structured test execution automation, we compare \shortname{} against the closely related AutoDroid. The comparison emphasizes \shortname{}'s strengths in error correction (66\% higher), execution success rate (1.7\% higher), and execution efficiency (8.7s lower time required per step) compared to AutoDroid. Furthermore, its use of distilled models enables cost-effective and scalable deployment.

\begin{table}
    \captionsetup{aboveskip=-12pt, belowskip=0pt}
    \centering
    \small
    \resizebox{\columnwidth}{!}{%
    \begin{tabular}{lccccc}
        \hline
        \textbf{Technique} & \textbf{Model} & \makecell{\textbf{Test}\\\textbf{Execution}\\\textbf{Success Rate}} & \makecell{\textbf{Error Recovery}\\\textbf{Rate}} & \makecell{\textbf{Execution}\\\textbf{Time}\\\textbf{per Step}} \\
        \hline
        AutoDroid & GPT-4o & 71.3\% & 12\%  & 20.5s \\
        \hline
        \shortname{} & GPT-4o & \textbf{73\%} & \textbf{78\%} & 11.8s \\
        \shortname{} & LLaMA-3.1 8B (Base) & 60\% & 10\% & \textbf{11.0s} \\
        \shortname{} & LLaMA-3.1 8B (Fine-tuned) & 65\% & 55\% & \textbf{11.0s} \\
        \hline
    \end{tabular}%
    }
    \caption{Test Execution Performance Metrics for Various Techniques and Models}
    \label{tab:rq1_results}
\end{table}


\begin{figure}
    \captionsetup{aboveskip=0pt, belowskip=-12pt}
    \centering
    \includegraphics[width=0.45\textwidth]{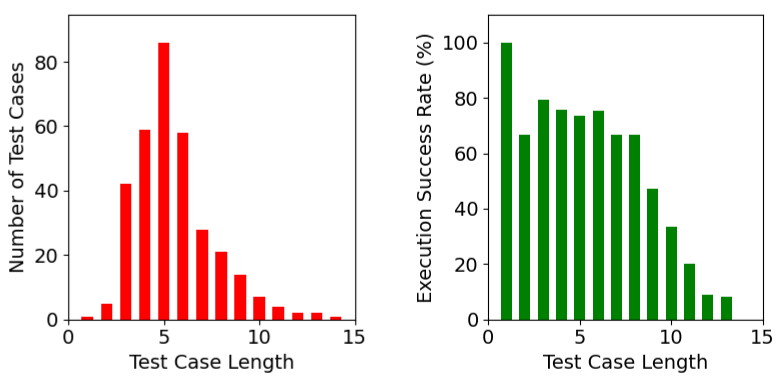}
    \caption{The left histogram shows test case length distribution, with most cases in the 3–10 step range. The right histogram shows that shorter test cases succeed more often, while longer ones have lower success due to increased interactions and error risks}
    \label{fig:rq1_length_success}
\end{figure}
\paragraph{Impact of Test Case Length on Execution Success}


Minimum number of test steps needed to complete a test case can affects execution success rate. Figure~\ref{fig:rq1_length_success} shows that most test cases have 3–10 steps (left) and that shorter test cases (fewer than 6 steps) succeed more often (right). Longer test cases see lower success due to increased interactions, error risks, and incorrect action selection. This highlights the need for refining test descriptions and improving error recovery.


\begin{table}[h]
    \captionsetup{aboveskip=-12pt, belowskip=0pt}
    \centering
    \begin{tabular}{|l|c|}
        \hline
        \textbf{Condition} & \textbf{Execution Success Rate} \\
        \hline
        Original Test Cases & 65\% \\
        Modified Test Cases & 73\% \\
        \hline
    \end{tabular}
    \caption{Impact of Test Case Modifications (Ambiguity Resolution of Test Case Descriptions) on Execution Success Rate}
    \label{tab:rq1_modification_results}
\end{table}

\paragraph{Impact of Ambiguity Resolution in Test Cases}

Ambiguous test descriptions reduced execution reliability. For example, \textit{"Change notification settings"} lacked details on which settings to modify, leading \shortname{} to misinterpret steps. Refining descriptions (e.g., \textit{"Disable email notifications from the Settings menu"}) improved success rates from 65\% to 73\% (Table~\ref{tab:rq1_modification_results}). Guidelines for resolving ambiguity are in Appendix~\ref{app:guide}.

\vspace{-0.5\baselineskip}
\section{Related Works}

Automated mobile application testing has traditionally relied on script-based frameworks such as \textit{Appium} and \textit{Espresso}~\cite{AppiumDocs, EspressoDocs}, which allow developers to write scripts that simulate user interactions. While these tools provide consistency in test execution, they require extensive manual effort to maintain scripts as applications evolve, making them inefficient for large-scale automation. To mitigate script maintenance overhead, record-and-replay mechanisms, such as \textit{RERAN}~\cite{lam2017record}, capture user interactions for deterministic execution. However, these approaches remain fragile when handling dynamic UI changes requiring manual adaptation of the scripts and limiting their effectiveness in real-world scenarios. In contrast, LELANTE's LLM driven execution directly interprets human-language test cases without requiring any pre-written scripts.



To improve scalability, search-based techniques such as \textit{Stoat} ~\cite{su2017guided} and \textit{Sapienz} ~\cite{mao2016sapienz} employ probabilistic and evolutionary algorithms for automated exploratory testing. Although effective in broadening test coverage, these methods often require substantial computational resources and manual validation of results. Recently, AI-driven methods such as \textit{AutoDroid}~\cite{wen2024autodroid} leverage LLMs to automate single-step UI interactions. However, AutoDroid primarily focuses on task execution without explicitly handling incorrect execution paths or error recovery. In contrast, \shortname{} adopts an iterative, reasoning-driven approach specifically designed for testers, emphasizing robustness through dynamic error detection and recovery.

Multimodal frameworks integrating vision-language models~\cite{liu2024vision, wu-etal-2024-mobilevlm} have also been explored for UI understanding and verification. However, these models typically involve high computational costs due to heavy reliance on visual processing, limiting practical deployment at scale. \shortname{}, on the other hand, relies solely on textual representation derived from a GUI refinement process (\S~\ref{sec:gui_refinement}), which enhances the effectiveness of prompt engineering, improves model performance, and reduces deployment costs.

\vspace{-0.5\baselineskip}
\section{Discussion}

\shortname{} demonstrates a significant step in automating Android testing, achieving a 73\% success rate and 78\% error recovery. However, performance declines with longer test cases, highlighting challenges in handling ambiguity and error propagation. While GUI refinement improves execution, reliance on manual test case modifications suggests limitations in natural language understanding. Another limitation of~\shortname{} is its inability to autonomously verify if an executed test case successfully achieves its expected results. Consequently, human validation remains essential to confirm whether the expected final application state is correctly attained. Additionally, the current backtracking mechanism remains relatively slow during error recovery, necessitating improvements on reasoning mechanism and disambiguation strategy.
\vspace{-0.5\baselineskip}
\section{Conclusion and Future Work}
In this work, we propose~\shortname{} , an LLM-driven framework for automating the execution of test cases for Android application. \shortname{} utilizes a refined screen representation and a structured prompt to guide an LLM in generating execution plans for test cases.  ~\shortname{}'s reasoning-driven action selection and automated backtracking for error correction results in an enhanced execution efficiency and adaptability to dynamic UI changes. Our work paves the way for further advancements in AI-driven test automation, reducing developer effort, improving software quality and eliminating the need for manual script writing for test case automation. Future efforts will focus on automating test outcome (i.e., expected results) validation to reduce human dependency, optimizing backtracking performance, and investigating prompt-based or incremental learning techniques for leveraging historical execution data without extensive model fine-tuning.

\vspace{-0.5\baselineskip}
\section{Acknowledgments}

This work was supported by Samsung R\&D Institute Bangladesh (SRBD) and Bangladesh University of Engineering and Technology (BUET).
\vspace{-0.5\baselineskip}
\bibliographystyle{ACM-Reference-Format}
\bibliography{custom}


\begin{thebibliography}{31}


\ifx \showCODEN    \undefined \def \showCODEN     #1{\unskip}     \fi
\ifx \showISBNx    \undefined \def \showISBNx     #1{\unskip}     \fi
\ifx \showISBNxiii \undefined \def \showISBNxiii  #1{\unskip}     \fi
\ifx \showISSN     \undefined \def \showISSN      #1{\unskip}     \fi
\ifx \showLCCN     \undefined \def \showLCCN      #1{\unskip}     \fi
\ifx \shownote     \undefined \def \shownote      #1{#1}          \fi
\ifx \showarticletitle \undefined \def \showarticletitle #1{#1}   \fi
\ifx \showURL      \undefined \def \showURL       {\relax}        \fi
\providecommand\bibfield[2]{#2}
\providecommand\bibinfo[2]{#2}
\providecommand\natexlab[1]{#1}
\providecommand\showeprint[2][]{arXiv:#2}

\bibitem[Achiam et~al\mbox{.}(2023)]%
        {achiam2023gpt}
\bibfield{author}{\bibinfo{person}{Josh Achiam}, \bibinfo{person}{Steven Adler}, \bibinfo{person}{Sandhini Agarwal}, \bibinfo{person}{Lama Ahmad}, \bibinfo{person}{Ilge Akkaya}, \bibinfo{person}{Florencia~Leoni Aleman}, \bibinfo{person}{Diogo Almeida}, \bibinfo{person}{Janko Altenschmidt}, \bibinfo{person}{Sam Altman}, \bibinfo{person}{Shyamal Anadkat}, {et~al\mbox{.}}} \bibinfo{year}{2023}\natexlab{}.
\newblock \showarticletitle{Gpt-4 technical report}.
\newblock \bibinfo{journal}{\emph{arXiv preprint arXiv:2303.08774}} (\bibinfo{year}{2023}).
\newblock


\bibitem[Coppola et~al\mbox{.}(2020)]%
        {coppola2020mobile}
\bibfield{author}{\bibinfo{person}{Riccardo Coppola}, \bibinfo{person}{Luca Ardito}, \bibinfo{person}{Maurizio Morisio}, {and} \bibinfo{person}{Marco Torchiano}.} \bibinfo{year}{2020}\natexlab{}.
\newblock \showarticletitle{Mobile testing: new challenges and perceived difficulties from developers of the Italian industry}.
\newblock \bibinfo{journal}{\emph{IT Professional}} \bibinfo{volume}{22}, \bibinfo{number}{5} (\bibinfo{year}{2020}), \bibinfo{pages}{32--39}.
\newblock


\bibitem[Developers(2025)]%
        {EspressoDocs}
\bibfield{author}{\bibinfo{person}{Android Developers}.} \bibinfo{year}{2025}\natexlab{}.
\newblock \bibinfo{title}{Espresso Testing Framework - Android Developers}.
\newblock
\urldef\tempurl%
\url{https://developer.android.com/training/testing/espresso}
\showURL{%
\tempurl}
\newblock
\shownote{Accessed: 2025-03-03}.


\bibitem[Dini et~al\mbox{.}(2018)]%
        {dini2018risk}
\bibfield{author}{\bibinfo{person}{Gianluca Dini}, \bibinfo{person}{Fabio Martinelli}, \bibinfo{person}{Ilaria Matteucci}, \bibinfo{person}{Marinella Petrocchi}, \bibinfo{person}{Andrea Saracino}, {and} \bibinfo{person}{Daniele Sgandurra}.} \bibinfo{year}{2018}\natexlab{}.
\newblock \showarticletitle{Risk analysis of Android applications: A user-centric solution}.
\newblock \bibinfo{journal}{\emph{Future Generation Computer Systems}}  \bibinfo{volume}{80} (\bibinfo{year}{2018}), \bibinfo{pages}{505--518}.
\newblock


\bibitem[Dosovitskiy et~al\mbox{.}(2021)]%
        {dosovitskiy2021an}
\bibfield{author}{\bibinfo{person}{Alexey Dosovitskiy}, \bibinfo{person}{Lucas Beyer}, \bibinfo{person}{Alexander Kolesnikov}, \bibinfo{person}{Dirk Weissenborn}, \bibinfo{person}{Xiaohua Zhai}, \bibinfo{person}{Thomas Unterthiner}, \bibinfo{person}{Mostafa Dehghani}, \bibinfo{person}{Matthias Minderer}, \bibinfo{person}{Georg Heigold}, \bibinfo{person}{Sylvain Gelly}, \bibinfo{person}{Jakob Uszkoreit}, {and} \bibinfo{person}{Neil Houlsby}.} \bibinfo{year}{2021}\natexlab{}.
\newblock \showarticletitle{An Image is Worth 16x16 Words: Transformers for Image Recognition at Scale}. In \bibinfo{booktitle}{\emph{International Conference on Learning Representations}}.
\newblock
\urldef\tempurl%
\url{https://openreview.net/forum?id=YicbFdNTTy}
\showURL{%
\tempurl}


\bibitem[Falaki et~al\mbox{.}(2010)]%
        {10.1145/1814433.1814453}
\bibfield{author}{\bibinfo{person}{Hossein Falaki}, \bibinfo{person}{Ratul Mahajan}, \bibinfo{person}{Srikanth Kandula}, \bibinfo{person}{Dimitrios Lymberopoulos}, \bibinfo{person}{Ramesh Govindan}, {and} \bibinfo{person}{Deborah Estrin}.} \bibinfo{year}{2010}\natexlab{}.
\newblock \showarticletitle{Diversity in smartphone usage}. In \bibinfo{booktitle}{\emph{Proceedings of the 8th International Conference on Mobile Systems, Applications, and Services}} (San Francisco, California, USA) \emph{(\bibinfo{series}{MobiSys '10})}. \bibinfo{publisher}{Association for Computing Machinery}, \bibinfo{address}{New York, NY, USA}, \bibinfo{pages}{179–194}.
\newblock
\showISBNx{9781605589855}
\href{https://doi.org/10.1145/1814433.1814453}{doi:\nolinkurl{10.1145/1814433.1814453}}


\bibitem[Gou et~al\mbox{.}(2021)]%
        {gou2021knowledge}
\bibfield{author}{\bibinfo{person}{Jianping Gou}, \bibinfo{person}{Baosheng Yu}, \bibinfo{person}{Stephen~J Maybank}, {and} \bibinfo{person}{Dacheng Tao}.} \bibinfo{year}{2021}\natexlab{}.
\newblock \showarticletitle{Knowledge distillation: A survey}.
\newblock \bibinfo{journal}{\emph{International Journal of Computer Vision}} \bibinfo{volume}{129}, \bibinfo{number}{6} (\bibinfo{year}{2021}), \bibinfo{pages}{1789--1819}.
\newblock


\bibitem[Grano et~al\mbox{.}(2018)]%
        {grano2018exploring}
\bibfield{author}{\bibinfo{person}{Giovanni Grano}, \bibinfo{person}{Adelina Ciurumelea}, \bibinfo{person}{Sebastiano Panichella}, \bibinfo{person}{Fabio Palomba}, {and} \bibinfo{person}{Harald~C Gall}.} \bibinfo{year}{2018}\natexlab{}.
\newblock \showarticletitle{Exploring the integration of user feedback in automated testing of android applications}. In \bibinfo{booktitle}{\emph{2018 IEEE 25Th international conference on software analysis, evolution and reengineering (SANER)}}. IEEE, \bibinfo{pages}{72--83}.
\newblock


\bibitem[Guo et~al\mbox{.}(2019)]%
        {guo2019sara}
\bibfield{author}{\bibinfo{person}{Jiaqi Guo}, \bibinfo{person}{Shuyue Li}, \bibinfo{person}{Jian-Guang Lou}, \bibinfo{person}{Zijiang Yang}, {and} \bibinfo{person}{Ting Liu}.} \bibinfo{year}{2019}\natexlab{}.
\newblock \showarticletitle{Sara: self-replay augmented record and replay for android in industrial cases}. In \bibinfo{booktitle}{\emph{Proceedings of the 28th acm sigsoft international symposium on software testing and analysis}}. \bibinfo{pages}{90--100}.
\newblock


\bibitem[Hou et~al\mbox{.}(2024)]%
        {hou2024large}
\bibfield{author}{\bibinfo{person}{Xinyi Hou}, \bibinfo{person}{Yanjie Zhao}, \bibinfo{person}{Yue Liu}, \bibinfo{person}{Zhou Yang}, \bibinfo{person}{Kailong Wang}, \bibinfo{person}{Li Li}, \bibinfo{person}{Xiapu Luo}, \bibinfo{person}{David Lo}, \bibinfo{person}{John Grundy}, {and} \bibinfo{person}{Haoyu Wang}.} \bibinfo{year}{2024}\natexlab{}.
\newblock \showarticletitle{Large language models for software engineering: A systematic literature review}.
\newblock \bibinfo{journal}{\emph{ACM Transactions on Software Engineering and Methodology}} \bibinfo{volume}{33}, \bibinfo{number}{8} (\bibinfo{year}{2024}), \bibinfo{pages}{1--79}.
\newblock


\bibitem[Hu et~al\mbox{.}(2022)]%
        {hu2022lora}
\bibfield{author}{\bibinfo{person}{Edward~J Hu}, \bibinfo{person}{Yelong Shen}, \bibinfo{person}{Phillip Wallis}, \bibinfo{person}{Zeyuan Allen-Zhu}, \bibinfo{person}{Yuanzhi Li}, \bibinfo{person}{Shean Wang}, \bibinfo{person}{Lu Wang}, \bibinfo{person}{Weizhu Chen}, {et~al\mbox{.}}} \bibinfo{year}{2022}\natexlab{}.
\newblock \showarticletitle{Lora: Low-rank adaptation of large language models.}
\newblock \bibinfo{journal}{\emph{ICLR}} \bibinfo{volume}{1}, \bibinfo{number}{2} (\bibinfo{year}{2022}), \bibinfo{pages}{3}.
\newblock


\bibitem[Jaiswal(2018)]%
        {jaiswal2018android}
\bibfield{author}{\bibinfo{person}{Manishaben Jaiswal}.} \bibinfo{year}{2018}\natexlab{}.
\newblock \showarticletitle{Android the mobile operating system and architecture}.
\newblock \bibinfo{journal}{\emph{Manishaben Jaiswal," ANDROID THE MOBILE OPERATING SYSTEM AND ARCHITECTURE", International Journal of Creative Research Thoughts (IJCRT), ISSN}} (\bibinfo{year}{2018}), \bibinfo{pages}{2320--2882}.
\newblock


\bibitem[Kong et~al\mbox{.}(2018)]%
        {kong2018automated}
\bibfield{author}{\bibinfo{person}{Pingfan Kong}, \bibinfo{person}{Li Li}, \bibinfo{person}{Jun Gao}, \bibinfo{person}{Kui Liu}, \bibinfo{person}{Tegawend{\'e}~F Bissyand{\'e}}, {and} \bibinfo{person}{Jacques Klein}.} \bibinfo{year}{2018}\natexlab{}.
\newblock \showarticletitle{Automated testing of android apps: A systematic literature review}.
\newblock \bibinfo{journal}{\emph{IEEE Transactions on Reliability}} \bibinfo{volume}{68}, \bibinfo{number}{1} (\bibinfo{year}{2018}), \bibinfo{pages}{45--66}.
\newblock


\bibitem[Konuk et~al\mbox{.}(2024)]%
        {konuk2024evaluation}
\bibfield{author}{\bibinfo{person}{Metin Konuk}, \bibinfo{person}{Cem Baglum}, {and} \bibinfo{person}{Ugur Yayan}.} \bibinfo{year}{2024}\natexlab{}.
\newblock \showarticletitle{Evaluation of Large Language Models for Unit Test Generation}. In \bibinfo{booktitle}{\emph{2024 Innovations in Intelligent Systems and Applications Conference (ASYU)}}. IEEE, \bibinfo{pages}{1--6}.
\newblock


\bibitem[Lam et~al\mbox{.}(2017)]%
        {lam2017record}
\bibfield{author}{\bibinfo{person}{Wing Lam}, \bibinfo{person}{Zhengkai Wu}, \bibinfo{person}{Dengfeng Li}, \bibinfo{person}{Wenyu Wang}, \bibinfo{person}{Haibing Zheng}, \bibinfo{person}{Hui Luo}, \bibinfo{person}{Peng Yan}, \bibinfo{person}{Yuetang Deng}, {and} \bibinfo{person}{Tao Xie}.} \bibinfo{year}{2017}\natexlab{}.
\newblock \showarticletitle{Record and replay for android: Are we there yet in industrial cases?}. In \bibinfo{booktitle}{\emph{Proceedings of the 2017 11th joint meeting on foundations of software engineering}}. \bibinfo{pages}{854--859}.
\newblock


\bibitem[Li et~al\mbox{.}(2024)]%
        {li2024test}
\bibfield{author}{\bibinfo{person}{Youwei Li}, \bibinfo{person}{Yangyang Li}, {and} \bibinfo{person}{Yangzhao Yang}.} \bibinfo{year}{2024}\natexlab{}.
\newblock \showarticletitle{Test-Agent: A Multimodal App Automation Testing Framework Based on the Large Language Model}. In \bibinfo{booktitle}{\emph{2024 IEEE 4th International Conference on Digital Twins and Parallel Intelligence (DTPI)}}. IEEE, \bibinfo{pages}{609--614}.
\newblock


\bibitem[Lin et~al\mbox{.}(2020)]%
        {lin2020test}
\bibfield{author}{\bibinfo{person}{Jun-Wei Lin}, \bibinfo{person}{Navid Salehnamadi}, {and} \bibinfo{person}{Sam Malek}.} \bibinfo{year}{2020}\natexlab{}.
\newblock \showarticletitle{Test automation in open-source android apps: A large-scale empirical study}. In \bibinfo{booktitle}{\emph{Proceedings of the 35th IEEE/ACM International Conference on Automated Software Engineering}}. \bibinfo{pages}{1078--1089}.
\newblock


\bibitem[Liu et~al\mbox{.}(2024b)]%
        {liu2024your}
\bibfield{author}{\bibinfo{person}{Jiawei Liu}, \bibinfo{person}{Chunqiu~Steven Xia}, \bibinfo{person}{Yuyao Wang}, {and} \bibinfo{person}{Lingming Zhang}.} \bibinfo{year}{2024}\natexlab{b}.
\newblock \showarticletitle{Is your code generated by chatgpt really correct? rigorous evaluation of large language models for code generation}.
\newblock \bibinfo{journal}{\emph{Advances in Neural Information Processing Systems}}  \bibinfo{volume}{36} (\bibinfo{year}{2024}).
\newblock


\bibitem[Liu et~al\mbox{.}(2024a)]%
        {liu2024vision}
\bibfield{author}{\bibinfo{person}{Zhe Liu}, \bibinfo{person}{Cheng Li}, \bibinfo{person}{Chunyang Chen}, \bibinfo{person}{Junjie Wang}, \bibinfo{person}{Boyu Wu}, \bibinfo{person}{Yawen Wang}, \bibinfo{person}{Jun Hu}, {and} \bibinfo{person}{Qing Wang}.} \bibinfo{year}{2024}\natexlab{a}.
\newblock \showarticletitle{Vision-driven automated mobile gui testing via multimodal large language model}.
\newblock \bibinfo{journal}{\emph{arXiv preprint arXiv:2407.03037}} (\bibinfo{year}{2024}).
\newblock


\bibitem[Mao et~al\mbox{.}(2016)]%
        {mao2016sapienz}
\bibfield{author}{\bibinfo{person}{Ke Mao}, \bibinfo{person}{Mark Harman}, {and} \bibinfo{person}{Yue Jia}.} \bibinfo{year}{2016}\natexlab{}.
\newblock \showarticletitle{Sapienz: Multi-objective automated testing for android applications}. In \bibinfo{booktitle}{\emph{Proceedings of the 25th international symposium on software testing and analysis}}. \bibinfo{pages}{94--105}.
\newblock


\bibitem[Rashidi and Fung(2015)]%
        {rashidi2015survey}
\bibfield{author}{\bibinfo{person}{Bahman Rashidi} {and} \bibinfo{person}{Carol~J Fung}.} \bibinfo{year}{2015}\natexlab{}.
\newblock \showarticletitle{A Survey of Android Security Threats and Defenses.}
\newblock \bibinfo{journal}{\emph{J. Wirel. Mob. Networks Ubiquitous Comput. Dependable Appl.}} \bibinfo{volume}{6}, \bibinfo{number}{3} (\bibinfo{year}{2015}), \bibinfo{pages}{3--35}.
\newblock


\bibitem[Sahin et~al\mbox{.}(2019)]%
        {sahin2019randr}
\bibfield{author}{\bibinfo{person}{Onur Sahin}, \bibinfo{person}{Assel Aliyeva}, \bibinfo{person}{Hariharan Mathavan}, \bibinfo{person}{Ayse Coskun}, {and} \bibinfo{person}{Manuel Egele}.} \bibinfo{year}{2019}\natexlab{}.
\newblock \showarticletitle{Randr: Record and replay for android applications via targeted runtime instrumentation}. In \bibinfo{booktitle}{\emph{2019 34th IEEE/ACM International Conference on Automated Software Engineering (ASE)}}. IEEE, \bibinfo{pages}{128--138}.
\newblock


\bibitem[Said et~al\mbox{.}(2020)]%
        {said2020gui}
\bibfield{author}{\bibinfo{person}{Kabir~S Said}, \bibinfo{person}{Liming Nie}, \bibinfo{person}{Adekunle~A Ajibode}, {and} \bibinfo{person}{Xueyi Zhou}.} \bibinfo{year}{2020}\natexlab{}.
\newblock \showarticletitle{GUI testing for mobile applications: objectives, approaches and challenges}. In \bibinfo{booktitle}{\emph{Proceedings of the 12th Asia-Pacific Symposium on Internetware}}. \bibinfo{pages}{51--60}.
\newblock


\bibitem[Sarkar et~al\mbox{.}(2019)]%
        {sarkar2019android}
\bibfield{author}{\bibinfo{person}{Anirban Sarkar}, \bibinfo{person}{Ayush Goyal}, \bibinfo{person}{David Hicks}, \bibinfo{person}{Debadrita Sarkar}, {and} \bibinfo{person}{Saikat Hazra}.} \bibinfo{year}{2019}\natexlab{}.
\newblock \showarticletitle{Android application development: a brief overview of android platforms and evolution of security systems}. In \bibinfo{booktitle}{\emph{2019 Third International conference on I-SMAC (IoT in Social, Mobile, Analytics and Cloud)(I-SMAC)}}. IEEE, \bibinfo{pages}{73--79}.
\newblock


\bibitem[Su et~al\mbox{.}(2017)]%
        {su2017guided}
\bibfield{author}{\bibinfo{person}{Ting Su}, \bibinfo{person}{Guozhu Meng}, \bibinfo{person}{Yuting Chen}, \bibinfo{person}{Ke Wu}, \bibinfo{person}{Weiming Yang}, \bibinfo{person}{Yao Yao}, \bibinfo{person}{Geguang Pu}, \bibinfo{person}{Yang Liu}, {and} \bibinfo{person}{Zhendong Su}.} \bibinfo{year}{2017}\natexlab{}.
\newblock \showarticletitle{Guided, stochastic model-based GUI testing of Android apps}. In \bibinfo{booktitle}{\emph{Proceedings of the 2017 11th joint meeting on foundations of software engineering}}. \bibinfo{pages}{245--256}.
\newblock


\bibitem[Team(2025)]%
        {AppiumDocs}
\bibfield{author}{\bibinfo{person}{Appium Team}.} \bibinfo{year}{2025}\natexlab{}.
\newblock \bibinfo{title}{Appium Documentation}.
\newblock
\urldef\tempurl%
\url{https://appium.io/docs/en/latest/}
\showURL{%
\tempurl}
\newblock
\shownote{Accessed: 2025-03-03}.


\bibitem[Touvron et~al\mbox{.}(2023)]%
        {touvron2023llama}
\bibfield{author}{\bibinfo{person}{Hugo Touvron}, \bibinfo{person}{Thibaut Lavril}, \bibinfo{person}{Gautier Izacard}, \bibinfo{person}{Xavier Martinet}, \bibinfo{person}{Marie-Anne Lachaux}, \bibinfo{person}{Timoth{\'e}e Lacroix}, \bibinfo{person}{Baptiste Rozi{\`e}re}, \bibinfo{person}{Naman Goyal}, \bibinfo{person}{Eric Hambro}, \bibinfo{person}{Faisal Azhar}, {et~al\mbox{.}}} \bibinfo{year}{2023}\natexlab{}.
\newblock \showarticletitle{Llama: Open and efficient foundation language models}.
\newblock \bibinfo{journal}{\emph{arXiv preprint arXiv:2302.13971}} (\bibinfo{year}{2023}).
\newblock


\bibitem[Wei et~al\mbox{.}(2022)]%
        {wei2022chain}
\bibfield{author}{\bibinfo{person}{Jason Wei}, \bibinfo{person}{Xuezhi Wang}, \bibinfo{person}{Dale Schuurmans}, \bibinfo{person}{Maarten Bosma}, \bibinfo{person}{Fei Xia}, \bibinfo{person}{Ed Chi}, \bibinfo{person}{Quoc~V Le}, \bibinfo{person}{Denny Zhou}, {et~al\mbox{.}}} \bibinfo{year}{2022}\natexlab{}.
\newblock \showarticletitle{Chain-of-thought prompting elicits reasoning in large language models}.
\newblock \bibinfo{journal}{\emph{Advances in neural information processing systems}}  \bibinfo{volume}{35} (\bibinfo{year}{2022}), \bibinfo{pages}{24824--24837}.
\newblock


\bibitem[Wen et~al\mbox{.}(2024)]%
        {wen2024autodroid}
\bibfield{author}{\bibinfo{person}{Hao Wen}, \bibinfo{person}{Yuanchun Li}, \bibinfo{person}{Guohong Liu}, \bibinfo{person}{Shanhui Zhao}, \bibinfo{person}{Tao Yu}, \bibinfo{person}{Toby Jia-Jun Li}, \bibinfo{person}{Shiqi Jiang}, \bibinfo{person}{Yunhao Liu}, \bibinfo{person}{Yaqin Zhang}, {and} \bibinfo{person}{Yunxin Liu}.} \bibinfo{year}{2024}\natexlab{}.
\newblock \showarticletitle{Autodroid: Llm-powered task automation in android}. In \bibinfo{booktitle}{\emph{Proceedings of the 30th Annual International Conference on Mobile Computing and Networking}}. \bibinfo{pages}{543--557}.
\newblock


\bibitem[Wu et~al\mbox{.}(2024)]%
        {wu-etal-2024-mobilevlm}
\bibfield{author}{\bibinfo{person}{Qinzhuo Wu}, \bibinfo{person}{Weikai Xu}, \bibinfo{person}{Wei Liu}, \bibinfo{person}{Tao Tan}, \bibinfo{person}{Liujian Liujianfeng}, \bibinfo{person}{Ang Li}, \bibinfo{person}{Jian Luan}, \bibinfo{person}{Bin Wang}, {and} \bibinfo{person}{Shuo Shang}.} \bibinfo{year}{2024}\natexlab{}.
\newblock \showarticletitle{{M}obile{VLM}: A Vision-Language Model for Better Intra- and Inter-{UI} Understanding}. In \bibinfo{booktitle}{\emph{Findings of the Association for Computational Linguistics: EMNLP 2024}}, \bibfield{editor}{\bibinfo{person}{Yaser Al-Onaizan}, \bibinfo{person}{Mohit Bansal}, {and} \bibinfo{person}{Yun-Nung Chen}} (Eds.). \bibinfo{publisher}{Association for Computational Linguistics}, \bibinfo{address}{Miami, Florida, USA}, \bibinfo{pages}{10231--10251}.
\newblock
\href{https://doi.org/10.18653/v1/2024.findings-emnlp.599}{doi:\nolinkurl{10.18653/v1/2024.findings-emnlp.599}}


\bibitem[Zimmermann and Koziolek(2023)]%
        {zimmermann2023gui}
\bibfield{author}{\bibinfo{person}{Daniel Zimmermann} {and} \bibinfo{person}{Anne Koziolek}.} \bibinfo{year}{2023}\natexlab{}.
\newblock \showarticletitle{GUI-Based Software Testing: An Automated Approach Using GPT-4 and Selenium WebDriver}. In \bibinfo{booktitle}{\emph{2023 38th IEEE/ACM International Conference on Automated Software Engineering Workshops (ASEW)}}. IEEE, \bibinfo{pages}{171--174}.
\newblock


\end{thebibliography}

\appendix
\vspace{-0.5\baselineskip}
\section{Guidelines for Formatting Test Cases} \label{app:guide}
Creating well-structured \textbf{test case descriptions} is crucial for ensuring the optimal performance of \shortname{}'s LLM-based processing pipeline. Poorly written test cases can introduce \textbf{ambiguity}, making it difficult for the model to interpret and execute the correct actions. For example, instead of saying \textit{"Add category inside the Edit Category window"}, a clearer version would be \textit{"Add a new category named \texttt{Elysium} from inside the Edit Category window"}. 


Additionally, test cases should be structured to provide only \textbf{essential details} while maintaining \textbf{precision}. If an action can be performed in multiple ways, only \textbf{one method} should be specified, such as \textit{"Set call vibration as \texttt{Basic call} in Call alerts and ringtones"} instead of just \textit{"Set call vibration as \texttt{Basic call}"}. \textbf{Unnecessary details} should be avoided when there is only \textbf{one way} to complete a task. When necessary, \textbf{explicit directions} should be provided to help locate options within the UI, such as clarifying that \textit{"Gallery settings are located in the drawer"}. Finally, test case descriptions should always include the \textbf{final confirmation step} to prevent premature completion—an instruction like \textit{"Update Samsung account profile picture by taking a picture"} should be revised to \textit{"Update Samsung account profile picture by taking a picture. Click \texttt{Done} to save"}.

\vspace{-0.5\baselineskip}
\section{Prompt structure} \label{app:prompt}

To ensure accurate test execution, \shortname{} employs a structured prompt shown below that guides the LLM in understanding the test case, evaluating prior actions, and determining the next step.
\vspace{-0.5\baselineskip}
\colorlet{shadecolor}{shadecolor2}
\begin{shaded} 
\ttfamily\scriptsize
\begin{lstlisting}
You are given the textual description of a screen, past actions, and an overall goal. Some actions have already been taken in the past and you need to determine the immediate next action needed to achieve the goal. Do the past actions indicate that the goal has already been achieved? Output a JSON with keys:
"goal_action_plan": "string, a goal is made up of a series of actions that need to be taken to achieve it. Write a detailed description of the actions required. Some actions may already have been completed. The next action may not be directly on the screen, but an action on the screen could lead to it. Think strategically, step by step, considering the goal and past actions.",
"past_actions_summary": "string, a short description of what actions were taken in the past",
"no_further_action_needed": "string, whether the past actions fully achieve the goal or not. Start with 'Past Actions indicate ...' or 'Past Actions do not indicate ...'. Don't forget to save if instructed.", "no_further_action_needed_bool": "boolean, true or false",
"immediate_next_action": "string, a detailed explanation of the next action to perform, based on the goal and past actions. Think step by step.",
"current_screen_actions": [ [action: str, ID: int] ],
"selected_current_screen_action": [reasoning: str, action: str, ID: int],
"repeating_past_action": "string, describing if a past action is being repeated, excluding scrolling or going back.",
"repeating_past_action_bool": "boolean, true or false. False for scrolling or going back.",
"id": "number/integer, the ID of the element to interact with. Use -1 if no action is needed.",
"text_input_value": "string, only EditText and AutoCompleteTextView can accept textinputvalue. For other elements, use '<NOVALUE>'"

\end{lstlisting}
Current Screen: \textbf{<Refined Current Screen Description>} \\
Overall Goal: \textbf{<Test Case Description>}
Past Actions:  \textbf{<Past Actions Selected by \shortname{}>} 
\end{shaded}










\end{document}